# Broadband high-resolution two-photon spectroscopy with laser frequency combs

Arthur Hipke[1,2], Samuel A. Meek[1], Takuro Ideguchi[1], Theodor W. Hänsch[1,2], Nathalie Picqué[1,2,3,*]

1. Max Planck Institut für Quantenoptik, Hans-Kopfermann-Str. 1, 85748 Garching, Germany
2. Ludwig-Maximilians-Universität München, Fakultät für Physik, Schellingstrasse 4/III, 80799 München, Germany
3. Institut des Sciences Moléculaires d'Orsay, CNRS, Bâtiment 350, Université Paris-Sud, 91405 Orsay, France
* nathalie.picque@mpq.mpg.de

***Abstract:*** *Two-photon excitation spectroscopy with broad spectral span is demonstrated at Doppler-limited resolution. We describe first Fourier transform two-photon spectroscopy of an atomic sample with two mode-locked laser oscillators in a dual-comb technique. Each transition is uniquely identified by the modulation imparted by the interfering comb excitations. The temporal modulation of the spontaneous two-photon fluorescence is monitored with a single photodetector, and the spectrum is revealed by a Fourier transform.*

Two-photon excitation is a widely-exploited nonlinear phenomenon with applications that extend to the fields of precision spectroscopy, photochemistry, and biochemical imaging. The increased availability of ultrashort-pulse lasers has been the trigger for a fast uptake of the two-photon excitation method in chemistry and biology, as they provide efficient excitation though usually do not allow for spectral discrimination. Here we demonstrate a new technique of two-photon excitation spectroscopy with two laser frequency combs that combines broad spectral bandwidth and Doppler-limited resolution.

High resolution may be achieved [1-5] in two-photon excitation with a train of pulses from a mode-locked laser, as the spectrum of an optical frequency comb is composed of evenly-spaced narrow lines. Even the first experiments with a picosecond laser demonstrated the potential of this technique for spectroscopy and spectral calibration [1,2]. The resonance condition for excitation of a given atomic energy level can be satisfied by many pairs of comb lines, so that the excitation probability of the level can be the same as for a resonantly tuned continuous-wave laser of the same average power. However, in this type of direct frequency comb spectroscopy, any energy level spacing can only be measured modulo the line spacing of the frequency comb, so that the technique is only suitable for very simple spectra. To disambiguate the signals from different energy levels, it has been proposed to modulate the output of a mode-locked laser with a varying arm Michelson interferometer [6]. Subsequently, a few investigations in the liquid phase at low resolution took advantage [7-9] of Michelson-based two-photon Fourier transform spectroscopy.

As already shown in linear absorption spectroscopy [10-14], two optical frequency comb generators of slightly different repetition frequencies also produce pairs of pulses (one from each comb) with a separation that changes linearly from pulse to pulse. They thus mimic a scanning Michelson interferometer, but, since there are no moving parts, the limitations in acquisition speed and resolution may be overcome.

We demonstrate such dual-comb two-photon spectroscopy with a proof-of-principle experiment on atomic rubidium vapor.

In dual-comb two-photon spectroscopy, two mode-locked femtosecond lasers with repetition frequencies $f_r$ and $f_r + \Delta f_r$, respectively, drive two-photon transitions in an atomic or molecular sample.

In the time domain, the excitation by pairs of pulses with a linearly increasing time separation produces Ramsey-like interference in the excitation amplitude. The population in the excited states depends on the competing processes of two-photon excitation, fluorescence decay and possibly other processes such as radiationless decay. Its modulations are recorded, as a function of time, via the fluorescence emitted during decays to lower states. The measured time-domain interference signal, the interferogram, is Fourier transformed to reveal the spectrum of the excited two-photon transitions.

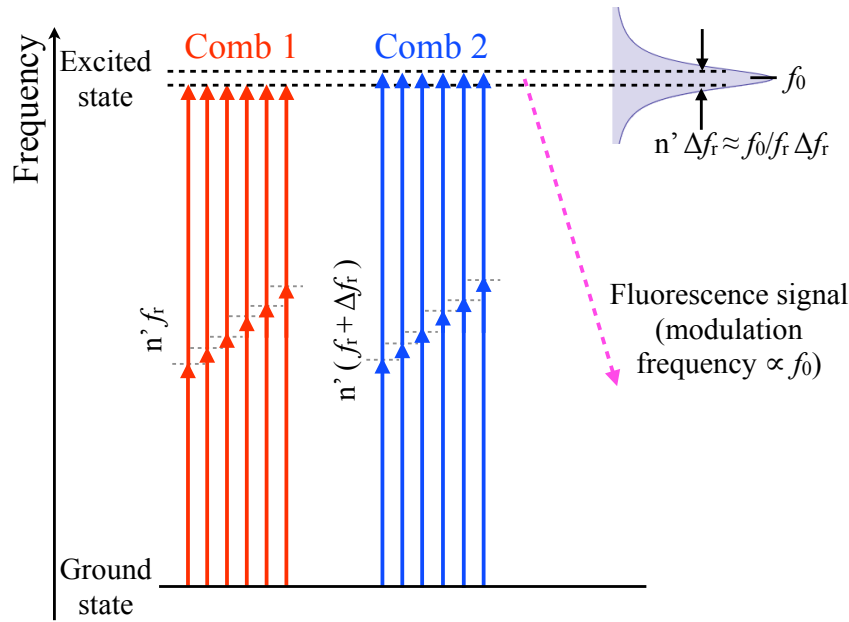

**Figure 1.** Frequency domain illustration of the principle of dual-comb two-photon spectroscopy. For simplicity, we assume in this figure that the carrier-envelope offset frequency of the two frequency combs is equal to zero. Pairs of comb modes with the same sum frequency (n' $f_r$ for Comb 1, n' $f_r + \Delta f_r$ for Comb 2) resonantly excite a two-photon transition at the optical frequency $f_0$. The excitation rate of the transition is modulated at the beat frequency n' $\Delta f_r$, which is approximately equal to $f_0 / f_r \, \Delta f_r$. An intensity modulation of frequency $f_0 / f_r \, \Delta f_r$ is thus observed in the fluorescence radiated during decays to lower states. As the frequency of the modulation is proportional to the transition frequency $f_0$, the two-photon excitation spectrum is revealed by Fourier transformation of the fluorescence intensity recorded versus time.

In the frequency domain, each laser produces a comb of modes with evenly-spaced optical frequencies. For instance, for Comb 1, the optical frequencies $f_{n,1}$ may be written $f_{n,1} = n f_r + f_{ceo,1}$, where n is an integer, and $f_{ceo,1}$ is the carrier-envelope offset frequency. Many pairs of modes (Fig.1) may satisfy the resonance condition for two-photon excitation at the same time. Possible sum frequencies, for Comb 1, are given by $f_{sum,1} = n' f_r + 2 f_{ceo,1}$ where n' is integer-valued. Two sum frequencies, one from each comb, interfere and the two-photon excitation rate is modulated at the beat note frequency n' $\Delta f_r + 2 \Delta f_{ceo}$, where $\Delta f_{ceo}$ is the difference of the carrier-envelope offset frequency of the two combs. The excited state fluorescence, which may be monitored through some indirect decay channel, is therefore modulated at the beat note frequencies within the excited line profile. The two-photon spectrum is thus mapped to radio-frequency beat notes that are directly proportional to the two-photon transition frequencies. The measurement of the repetition frequency and carrier-envelope offset frequency of the combs allows for converting the measured radio-frequencies to the proper optical scale.

For a first proof-of-principle experiment, we have chosen two-photon spectroscopy in atomic rubidium vapor with a resonant intermediate state. This level can be resonantly excited by one photon and its excitation rate is modulated by the interference of the exciting comb lines. The resulting population pulsations can be detected in the fluorescence light after stepwise two-photon excitation. Thus the fluorescence light is modulated at three groups of beat frequencies: *i)* m $\Delta f_r + \Delta f_{ceo}$, corresponding to one-photon comb excitation from the ground to the intermediate state, *ii)* k $\Delta f_r + \Delta f_{ceo}$, corresponding to one-photon comb excitation from the intermediate to the final excited state and *iii)* n' $\Delta f_r + 2 \Delta f_{ceo}$, corresponding to two-photon comb excitation from the ground to the final excited state. The Fourier transform of the interferogram reveals a spectrum mapping the one- and two-photon transitions in unequivocal regions.

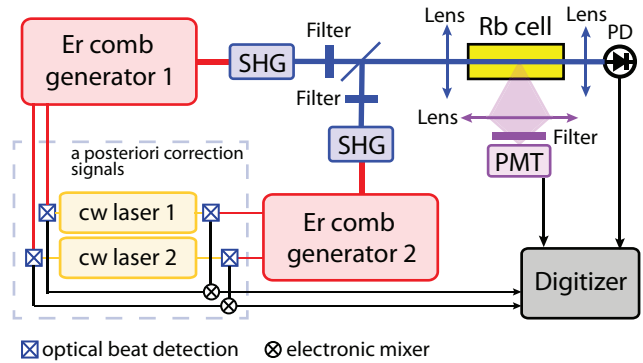

**Figure 2.** Experimental setup for dual-comb two-photon spectroscopy of atomic rubidium vapor. cw, continuous-wave; SHG, second-harmonic generation; PMT, photomultiplier tube; PD, photodiode.

The experimental setup is shown in Fig.2. Two amplified erbium-doped fiber comb generators with a repetition frequency of about 100 MHz emit pulses of a duration of the order of 100 fs. Their emission is centered around 192 THz (1.56 $\mu$m) and their average power is 150 mW. To prevent long term drifts, the repetition frequency of each comb is stabilized (not shown in the figure) against a radio-frequency clock via a mirror of the laser cavity mounted onto a piezoelectric transducer. One optical line of each comb is phase-locked to a free-running continuous-wave erbium-doped fiber laser emitting at 192 THz (1.56 $\mu$m) by feedback on the current of the pump diodes of the combs. The outputs of the erbium femtosecond lasers are frequency-doubled in magnesium-oxide doped periodically-poled lithium niobate (MgO:PPLN) crystals of a thickness of 0.5 mm, yielding spectra centered around 383 THz (783 nm), with a full-width at half-maximum of about 10 THz (20 nm). Dichroic mirrors isolate the second-harmonic radiation. The two 383-THz beams are then combined on a beam-splitter cube and focused into a heated 10-mm long cell filled with rubidium in natural abundance. Each frequency-doubled comb has an average power of about 20 mW at the sample. The fluorescence of the sample is observed through a side-window. It is filtered with an interference filter to reduce scattered light from the laser beam and detected with a photomultiplier tube. The laser beams that are transmitted through the sample may meanwhile beat on a fast Si photodiode, thus allowing for measuring the dual-comb linear absorption interferogram. The outputs of the detectors are digitized with a 14-bit analog-to-digital converter at a sampling rate up to 125 MS/s.

As already discussed e.g. in [13], stabilizing the combs to a radio-frequency reference involves low-bandwidth feedback loops that do not compensate for the short-term instabilities of the combs. Such instabilities severely distort the spectra. We therefore monitor the relative fluctuations of the two combs and apply an *a posteriori* correction algorithm similar to that reported in [14]. The beat signals between two pairs of individual comb lines of the two different combs in two different spectral regions are isolated with the aid of two auxiliary continuous-wave erbium fiber lasers emitting at 192 THz (1.56 $\mu$m) and 195 THz (1.53 $\mu$m), respectively, as local oscillators. Electronic mixers cancel out any fluctuation of the local oscillators. The two resulting radio-frequency beat signals are recorded, as a function of time, simultaneously to the interferogram. They provide the necessary correction signals of the timing jitter and relative fluctuations of the carrier-envelope phase of the combs. They are used in a computer algorithm to multiply the interferogram by a time-varying phase factor and to subsequently correct for timing fluctuations by adjusting the sampling rate to an equidistant grid. The correction signals are derived from the outputs of the erbium femtosecond oscillators around 192 THz (fundamental frequency). In order to properly correct the one-photon spectra that are centered around 384 THz (second harmonic) and the two-photon spectra that are centered around 768 THz (fourth harmonic), we multiply the interferometric phase by an appropriate order factor (2 for the second harmonic, 4 for the fourth harmonic). We thus extend previous correction schemes [14] to dual-comb nonlinear spectroscopy and show that monitoring the relative instabilities of the oscillators at their fundamental frequency is effective for correction of interferograms involving nonlinear frequency generation and/or nonlinear phenomena.

We illustrate Doppler-limited resolution with broadband one- and two-photon excitation spectra of $^{85}$Rb and $^{87}$Rb. The relevant energy levels and transitions of Rb are shown in Fig.3a. Within the excitation range of our lasers, rubidium has three one-photon resonances ($5S_{1/2}$-$5P_{3/2}$, $5P_{3/2}$-$5D_{3/2}$, and $5P_{3/2}$-$5D_{5/2}$) and two two-photon lines ($5S_{1/2}$-$5D_{3/2}$ and $5S_{1/2}$-$5D_{5/2}$). The $5S_{1/2}$ ground state of both isotopes is split (not shown in the figure) by a few GHz due to hyperfine interaction. The hyperfine splitting of the 5P and 5D states is narrower than the Doppler linewidths (about 1 GHz) and is thus not resolved in our spectra. From the 5D states, Rb atoms can decay into the 6P manifold, from where they emit fluorescence at 711 and 713 THz to repopulate the 5S ground state. We thus excite Rb into the 5D states with our dual-comb system and collect the resulting fluorescence to measure our interferograms. The intensity of the fluorescence is modulated at all beat note frequencies corresponding to the resonances excited by the combs.

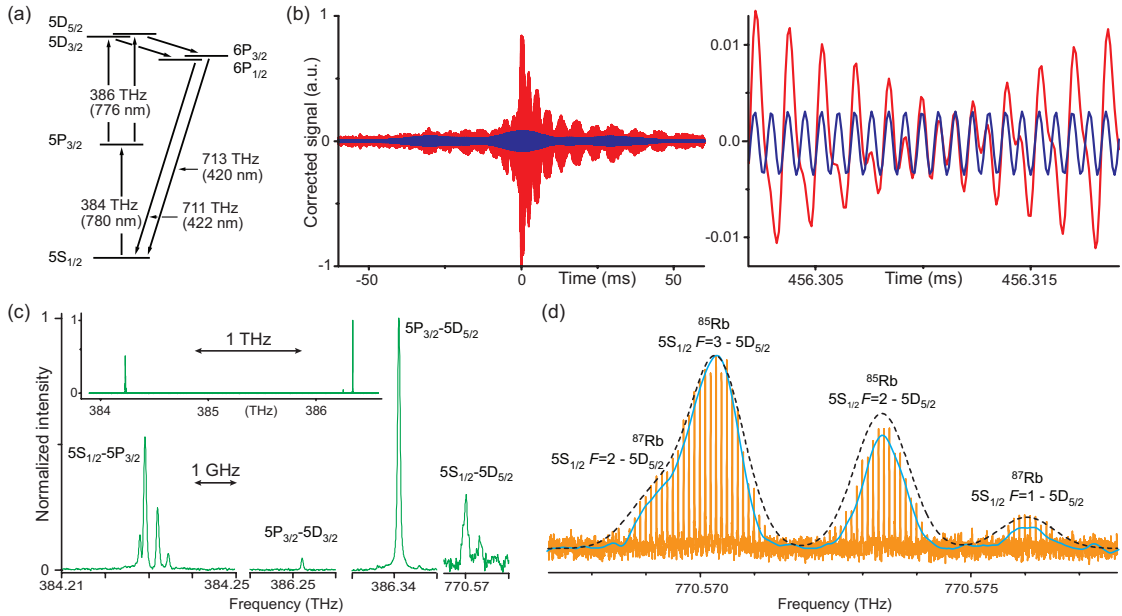

**Figure 3.** (a) Experimentally-relevant fine structure of rubidium. (b) Two portions of an unaveraged, corrected interferogram recorded at T=393 K with $\Delta f_r$=1 Hz (red trace) and contributions to that interferogram from two-photon transitions (blue trace). (c) Portions of the experimental dual-comb one- and two-photon excitation spectra of natural, atomic rubidium. The entire range of the one-photon excitation lines is shown in the upper left part of the panel, while the main part zooms into the individual one- and two-photon transitions (tick increment: 1 GHz). The spectrum, displayed at an unapodized resolution of 300 MHz, is recorded at T=308 K within 18-s effective measurement time. (d) Portion of the two-photon spectrum recorded at T=393 K (solid cyan: 18-s measurement time, unapodized resolution of 300 MHz; solid orange: 54-s measurement time, unapodized comb transform-limited linewidth of 4.4 MHz; dashed black: computed).

The relatively long lifetime of the 5D states (about 240 ns) acts as a low pass filter on the modulated fluorescence signal in the radio-frequency domain, limiting the maximum modulation frequency to a few MHz. Because of this limitation, we must choose a relatively low repetition frequency difference (here, $\Delta f_r$ =1 Hz).

While the entire spectral range of excitation simultaneously spans from 373 to 393 THz and from 755 to 780 THz (corresponding to 1.17 $10^5$ spectral elements, as defined by range of excitation/resolution), only four narrow frequency segments (Fig. 3c) exhibit spectral lines.

Figure 3d displays the frequency segment centered around 770.572 THz from a single interferogram, Fourier transformed at two different resolutions. The interferogram is recorded at high temperature (393 K), which is required to obtain a reasonable signal-to-noise ratio in the two-photon transitions. The spectra are calibrated using the measured repetition frequencies of the combs and one of the Rb lines. The spectrum of Fig. 3c is calibrated against the $^{85}$Rb $5S_{1/2}$, F=2 - $5P_{3/2}$ transition [15,16], whereas the spectra of Fig. 3d are calibrated against the $^{85}$Rb $5S_{1/2}$, F=2 - $5D_{5/2}$ transition [17]. The other line positions then agree within 40 MHz with the centroid values derived from refs. [15-17]. The two-photon lines are correctly reproduced by calculations that incorporate all allowed hyperfine transitions and assume Doppler profiles. For a measurement duration of 54 seconds (Fig. 3d), the spectrum is composed of individual comb lines, showing that the coherence between the two mode-locked lasers is reconstructed.

The above proof-of-principle demonstration shows the potential of dual-comb spectroscopy for the measurement of broad spectral bandwidth Doppler-limited two-photon excitation spectra. The technique applies to any sample that fluoresces under two-photon excitation and could be extended to multi-photon fluorescence or ionization. It permits highly-multiplexed measurements since all one- and two-photon transitions within the excitation bandwidth of the femtosecond lasers are simultaneously acquired within a short measurement time. Background-free fluorescence detection allows for higher sensitivity than absorption measurements.

Dual-comb two-photon spectroscopy could readily be applied for spatially selective real-time spectroscopy of liquid samples. As only a single photodetector is needed, it may easily be implemented in existing two-photon microscopes for hyperspectral mapping. In these applications, with the same experimental set-up, dual-comb two-photon spectroscopy can be complemented with e.g. dual-comb coherent anti-Stokes Raman spectroscopy [18] to simultaneously interrogate vibronic and vibrational transitions. Moreover, future work will aim at extending our technique to sub-Doppler resolution and hence to the field of precision spectroscopy. In contrast to direct frequency-comb two-photon spectroscopy with a single mode-locked laser, the optical free spectral range in dual-comb two-photon spectroscopy has no other limit than the spectral bandwidth of the laser frequency combs. Thus very high-resolution spectroscopy of highly crowded molecular spectra should become feasible.

Experimental support from Birgitta Bernhardt and Antonin Poisson is warmly acknowledged. Support by the European Research Council (Advanced Investigator Grant 267854), the Max Planck Foundation, the European Laboratory for Frequency Comb Spectroscopy and the Munich Center for Advanced Photonics are acknowledged.